# Exploring Jukebox: A Novel Audio Representation for Music Genre Identification in MIR

**Navin Kamuni [1], Mayank Jindal [2], Arpita Soni [3], Sukender Reddy Mallreddy [4], Sharath Chandra Macha [5]**

[1]navin.kamuni@gmail.com, [2]mayank.jindal5@gmail.com, [3]soni.arpita@gmail.com,
[4]sukender@ieee.org, [5]macha.sharathchandra@gmail.com

[1] AI-ML, BITS Pilani WILP, USA
[2] Independent Researcher, USA
[3] Eudoxia Research Centre, USA
[4] City of Dallas, USA
[5] CBRE Inc, USA

*Abstract*— For Music Information Retrieval (MIR) downstream tasks, the most common audio representation is time-frequency-based, such as Mel spectrograms. In order to identify musical genres, this study explores the possibilities of a new form of audio representation—one of the most usual MIR downstream tasks. Therefore, to discretely encoding music using deep vector quantization; a novel audio representation was created for the innovative generative music model i.e. Jukebox. The effectiveness of Jukebox's audio representation is compared to Mel spectrograms using a dataset that is almost equivalent to State-of-the-Art (SOTA) and an almost same transformer design. The results of this study imply that, at least when the transformers are pretrained using a very modest dataset of 20k tracks, Jukebox's audio representation is not superior to Mel spectrograms. This could be explained by the fact that Jukebox's audio representation does not sufficiently take into account the peculiarities of human hearing perception. On the other hand, Mel spectrograms are specifically created with the human auditory sense in mind.

## I. INTRODUCTION

This study explores the field of Music Information Retrieval (MIR), with a particular emphasis on music genre detection in order to compare a new approach to audio representation with the well-established use of Mel spectrograms. Waveform encoding of music has historically presented considerable hurdles for Machine Learning (ML) [1]–[6] because of its long-term dependencies; a 30-second—quality audio sample has more than 1 million time steps. As a result, spectrograms have become indispensable in MIR, converting audio inputs into a frequency representation that greatly shortens sequence lengths and thereby simplifies these complications.

However, deep Vector Quantization (VQ) [7], a novel audio representation presented by the generative music model Jukebox, emerges against these limitations. The ability to create believable music directly in the raw audio domain using deep VQ is a significant advancement and highlights the challenges that come with audio's long-term dependencies. It does this by compressing audio using a Neural Network (NN) to create a fixed-sized set of vectors, or codebooks, where each vector is indexed and represented by a token. This makes it possible to express audio as a two-dimensional codebook vector series or as a one-dimensional token sequence. Though the 2D codebook representation appears to be less structured than spectrograms, Jukebox's successful use of it to generate music points to a potential NN application. Therefore, the main goal of this study is to determine which method performs better in music genre recognition: Mel spectrograms or deep VQ. By training transformer using Mel spectrograms, deep VQ tokens, or deep VQ codebooks, this comparison is made operational. The reasoning for this investigation is based on the theory that an auditory model that can produce music—a sign of a more profound comprehension of its composition—might perform well in MIR challenges. This concept is consistent with the idea that the process of creation contributes to a better understanding of music, as demonstrated by Deep VQ's powers in music generation—an area in which spectrograms are yet to produce compelling results. Beyond philosophical reasons, deep VQ's technological prowess—particularly its remarkable compression capabilities—is the driving force behind its exploration. In comparison to the modern industry standard audio codec Opus [8], deep VQ tokens show better compression efficiency and provide a more simplified musical representation than spectrograms. This deep VQ feature has the potential to revolutionize music annotation and recommender systems by increasing efficiency and accuracy in MIR tasks. Against the background of conventional spectrograms, this study concludes by critically analyzing the potential of deep VQ as an improved audio representation for music genre recognition. The study intends to clarify whether deep VQ's new approach to audio encoding can outperform traditional methods, indicating its feasibility for wider applications in MIR, by using music genre recognition as a benchmark. With broad implications for improving our relationship with and comprehension of music through technological improvements, this investigation not only questions accepted paradigms in audio representation but also opens up new directions for study and growth in the field.

This paper is as follows; related works are shown in the following section. The dataset analysis is provided in Section III. The experimental setup and materials and procedures are covered in Section IV. The outcome analysis is provided in Section V. The study's limitations are described in Section VI, and Section VII offers some conclusions and ideas for further research.

## II. RELATED WORKS

In this study, we examine how deep learning models can be used to process audio representations for music genre recognition. Token- and CodebookFormer [9] are used to process deep VQ based audio representations, while SpectroFormer [10] is used to process Mel spectrograms. This section explores the theoretical foundations of time-frequency audio representations such as Mel spectrograms, as well as the foundational research papers supporting these models and the technologies enabling their functions, such as Jukebox and VQ-VAE. Presented by [11] in 2020, Jukebox is a unique deep learning model that can produce music in the unprocessed audio domain. It takes advantage of a new type of audio representation called deep VQ to overcome the difficulties caused by the long-range dependencies that are associated with audio sequences. A 30-second audio clip at a standard sampling rate of 44,100 Hz has a sequence length of 1.3 million, which poses a significant challenge to current models. Jukebox gets around this by using deep VQ compression and a hierarchical representation of the audio, notably by using the VQ-VAE. In order to fully explore the genre recognition potential of this study, it is imperative that music must be compressed while maintaining quality. This approach allows audio to be represented at various temporal resolutions. In order to model the latent space with a distribution of codebook vectors, each represented by a symbolic token. Compared to conventional VAEs, this representation produces a greater compression rate, suggesting an alternative approach for effective music genre detection.

The study also discusses Wavenet, a NN that [12] created in 2016 to produce raw audio signals, such as instrumental music and voices that resemble people. Jukebox's VQ-VAE was inspired to apply similar techniques by Wavenet's use of dilated convolutions to extend the receptive field of convolution kernels, which represents a substantial development in modelling long-term relationships in raw audio. Another key model impacting this study is Bidirectional Encoder Representations from Transformers (BERT), which was created by [13] in 2018. The bidirectional strategy used by the models in this study for music genre categorization is inspired by BERT's bidirectional training of natural language representations [14]–[18], which enhances performance by taking into account both the left and right contexts for word prediction. MusiCoder, created by [19] in 2021, uses elements from audio preprocessing methods like Mel spectrograms to apply BERT's principles to music genre detection. Inspired by MusiCoder, the SpectroFormer model in this study investigates the efficacy of Mel spectrograms in genre classification. Mel spectrograms, which represent the subtle aspects of sound as experienced by humans, are made by computing short-time Fourier transformation spectrograms, mapping frequencies to Mel scale, and then converting amplitude to decibels. This study also makes reference to [20] work, which highlights the adaptability and promise of VQ-VAE-based techniques in MIR problems to compress Mel spectrograms for emotion and theme recognition in music. This study's examination of music genre recognition has its foundation in the developments and approaches of multiple significant research publications and models. These technologies, which range from the generative powers of Wavenet and Jukebox to the representational effectiveness of BERT and MusiCoder, as well as the perceptual alignment of Mel spectrograms, all contribute to the theoretical underpinnings and experimental methodology of the deep learning models used in this study.

## III. DATASET ANALYSIS

In order to assess different audio representations, this study investigates the genre recognition of music, using the Free Music Archive (FMA) dataset as a key resource [11]. The FMA offers a comprehensive library of 106,574 recordings by 16,341 artists over 161 genres, and is curated by WFMU, America's longest-standing freeform radio station. It stands out for its absence of commercial influences and genre limits. The FMA dataset is especially useful for tasks involving genre classification since the tracks, which are recorded in mp3 format and have an average bit rate of 263 kbit/s and a sampling rate of 44,100 Hz, are arranged in hierarchical genre taxonomy, selected by the artists themselves. There are four sizes for the FMA dataset: small (8,000 tracks), medium (25,000 tracks), large (106,574 tracks), and full (106,574 tracks). The first three sizes include 30-second playbacks of each audio. The medium-sized dataset is chosen for its suitability in providing a significant yet feasible amount of data, preventing the significant data engineering demands of the full-sized dataset, with the study focusing on using genre recognition for comparisons rather than accomplishing breakthrough findings. While the hierarchical genre taxonomy of the dataset makes classification jobs easier, the study chooses top-level genre categorizing for manageability and generalizability, particularly as certain genres are underrepresented. Despite its imbalanced genre distribution, the medium-sized dataset was selected for experimentation because it offers a single top-level genre per track, making it appropriate for easier, single-genre classification tasks. The study follows the preset dataset split of the FMA, which allocates 10% for testing and validation and 80% for training, in order to ensure experimental consistency and repeatability. In addition to maintaining the distribution of genres among sets, this split guarantees that a given artist's recordings are only available in a single dataset segment, which is essential for preventing overly optimistic accuracy results.

This methodological decision is in line with industry best practices for managing uneven datasets and distribution of artists.

### A. Dataset Preprocessing

Since, the two types of audio have different features and needs, the preprocessing stage of this study is essential to transforming raw audio into Mel spectrograms or deep VQ audio representations. Different approaches are used for each of these two types of audio representations. To reduce conflicts caused by software package dependencies, the approach uses two distinct code repositories: one for deep VQ preprocessing and another for Mel spectrogram preparation and music labelling. Even though the Jukebox developers had pre-trained the deep VQ model, it still required a great deal of code exploration to be applied in inference. Jukebox's VQ Variational Auto Encoder (VQ-VAE) performs deep VQ preprocessing at 44.1 kHz sampling rate and 8x, 32x, and 128x compression rates (refer to Fig. 1). In order to meet the issue of managing long sequences in ML models, the greatest compression rate of 128x is chosen to minimize sequence length. An integer is used to represent each token in the sequence of compressed audio tokens, which have a vocabulary size of 2048 and are similar to letters in an alphabet. After going through this approach, the 400MB FMA medium-sized dataset becomes manageable, allowing for direct loading into RAM to expedite training. On the contrary hand, Mel spectrogram preprocessing makes use of the Python Librosa[1] package, and the resulting spectrograms are saved on the hard drive in order to prevent regeneration during every training epoch.

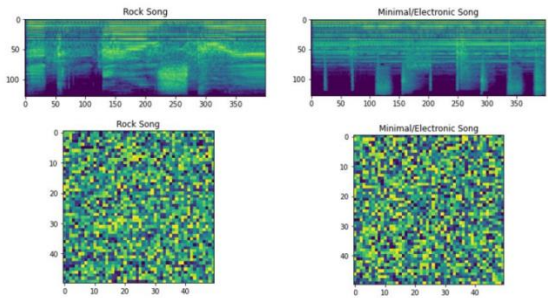

Fig. 1. A series of tokens (bottom) and spectrograms (top) depict a rock and an electronic/minimal tune. The VQ-VAE encrypts the token sequence. Every token has a distinct hue. The sequences are listed both top to bottom and left to right.

### IV. MATERIALS AND METHODS

In order to classify musical genres, this study looks into the possible benefits of using VQ-VAEs for audio preprocessing as opposed to more conventional techniques like Mel spectrograms. In order to investigate this research question, three distinct transformer-based models were compared in terms of their performance in genre classification: SpectroFormer (which receives Mel spectrograms as input), TokenFormer (which receives tokens generated by VQ-VAE), and CodebookFormer (which receives codebooks generated by VQ-VAE). Every model follows the same set of hyperparameters, which are shaped by the BERT and MusiCoder, for each model. But because the task is primarily audio-focused, the models have only four layers, which is a lower level of complexity than BERT and more in line with MusiCoder's architecture. Pretraining, which uses self-supervised techniques specific to each model's input type, is an essential part of model preparation. To improve its predictive power, TokenFormer is subjected to hidden Language Modelling, equivalent to BERT but with a larger proportion of tokens hidden. This method is imitated in CodebookFormer's pretraining, except relevant codebook entries are replaced with zero instead of the [MASK] token. In contrast, SpectroFormer's pretraining strategy uses static mask lengths and various masking percentages to adjust to its better temporal resolution while masking specific areas of the spectrogram. The different tasks of TokenFormer, CodebookFormer, and SpectroFormer are reflected in the choice of loss functions, which are cross-entropy for TokenFormer and CodebookFormer and Huber loss for SpectroFormer, respectively.

Using the FMA medium dataset—PyTorch Lightning[2], a high-level PyTorch API, is used in the experimental framework. Its purpose is to simplify the development process by keeping engineering chores and scientific code separate. This separation is a great option for the various training setups used in this research since it produces a number of advantages such as hardware-agnostic models, more readable code, and a lower frequency of defects. By using contemporary GPU scaling techniques, this configuration essentially doubles the amount of RAM that is available, giving the transformers access to up to 160 GB of RAM. This ability guarantees strong training performance, which is essential given the rigorous requirements of the models this study uses.

### A. Experimental Setup

This study's experimental design comprises three models—SpectroFormer, CodebookFormer, and TokenFormer—with six different configurations examined both with and without pretraining. Carefully choosing the hyperparameters to use during training is necessary to maximize RAM usage. This is because controlling the length of the input sequence affects RAM consumption, which in turn affects training efficacy. In order to improve genre classification performance, the training regimen uses the same hyperparameters for both the pretraining and regular classification phases according to [19]. The goal is to maximize the input sequence length while staying within the bounds of the available RAM. An important consideration for successful training results is the overall number of sequence components in each batch, which is also taken into account by this method. This study presents a difficulty in terms of data volume because the models are trained on a dataset of about 20k songs, whereas MusiCoder was pretrained on a substantially larger dataset of 152k songs. Therefore, a learning rate optimizes the training process by modulating the learning rate dependent on the training step, as implemented by [19]. Class weighting also corrects for the unequal distribution of genres in the FMA medium dataset, making minor classes fairly represented and mitigating model bias towards more common genres. When fine-tuning pretrained models, a static learning rate is used in conjunction with a thorough search throughout a predetermined parameter space to choose the best configuration based on validation set performance. To maintain consistency throughout the training process, this phase keeps the previously set training hyperparameters according to [19].

---

[1] https://librosa.org/doc/

[2] https://lightning-ai.webpkgcache.com/doc/-/s/lightning.ai/docs/pytorch/stable//index.html

However, in order to establish a fair comparison between the SpectroFormer and the VQ-VAE-based models (TokenFormer and CodebookFormer), the Mel spectrograms' hyperparameters are adjusted to match the tokens and codebooks produced by the VQ-VAE. We modify the spectrogram parameters to reflect the fact that the VQ-VAE uses fixed hop and frame sizes. In order to ensure consistency in temporal resolution, the hop size of the VQ-VAE is set at 128. This means that the Mel spectrograms produced by SpectroFormer must have the same value. In accordance with industry standards for audio signal processing, the frame size of Mel spectrograms is set as a power of two, ending at 512 even though the VQ-VAE has a frame size of 480. Given that SpectroFormer's frame size (11 ms) is substantially smaller than the standard recommendation for MIR tasks, this frame size disparity raises questions regarding the comparability of music genre classification performance. While this smaller frame size is beneficial for music production in VQ-VAE, it might not be the best for classifying genres, which could lead to unfair comparisons with models such as MusiCoder, which uses a 46ms frame size. The Mel spectrograms are processed using a Hann function in order to address the weighting of audio signals inside frames (refer to Fig. 2), a feature of the VQ-VAE attributed to its usage of residual connections. By smoothing the frames using a bell-shaped curve, this feature replicates the VQ-VAE's focus on mid-frame sounds and provides a fairer foundation for comparison. The frequency resolution, which is correlated with the frame size, determines the final hyperparameter, or the total number of Mel bands. Selecting a number of Mel bands equivalent to the frequency resolution guarantees that frequency information is not unnecessarily up-or down-scaled. SpectroFormer uses 86 Mel bands for its Mel spectrograms, which aligns the frequency resolution with the spectrogram's resolution requirements based on calculations made based on the selected frame size and sampling rate.

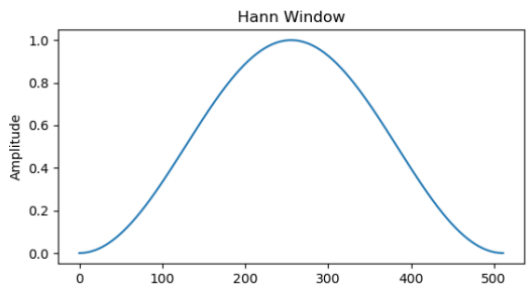

Fig. 2. The Mel Spectrogram frames are smoothed using the Hann function

## V. Result Analysis

Fig. 3 summarizes the F1 scores on the validation set for SpectroFormer, TokenFormer, and CodebookFormer in the comparative examination of music genre classification performance. A key metric for assessing the accuracy of the models is the F1 score, which is the harmonic mean of precision and recall. By computing the F1 score for each genre separately and then averaging them without weighting, the macro-averaged F1 score approach makes sure that each genre contributes equally to the final score. A model's random prediction would have a 1/16 chance of being true given that the objective entails categorizing among 16 genres, corresponding to an F1 score of 0.11. This sets the initial F1 score for the tests, and any model performance below this cutoff is regarded as being no more accurate than a random guess. As shown in Fig. 3, the finetuning arrangement with a learning rate of 2e-5 and a batch size of 16 proved to be the most successful of all the evaluated settings. All around, performance was somewhat worse with different finetuning options. All models perform better with pretrained versions, but SpectroFormer is significantly better at genre categorization, especially with pretrained versions. This result casts doubt on the study's original claim that token- or codebook-based deep VQ-based audio representations could perform better in genre classification tasks than conventional Mel spectrograms. In the context of the methodology and configurations evaluated, the data collected from the experimental setup of this work clearly contradicts this assumption, highlighting the higher efficacy of Mel spectrograms in music genre recognition. Based on the comparative study of macro-averaged F1 scores, the study concludes that Mel spectrograms are clearly preferred over deep VQ-based representations for music genre classification.

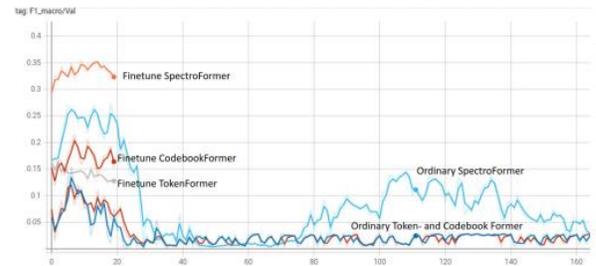

(a) Classification performance on the validation set. The y scale describes the macro F1 scale.

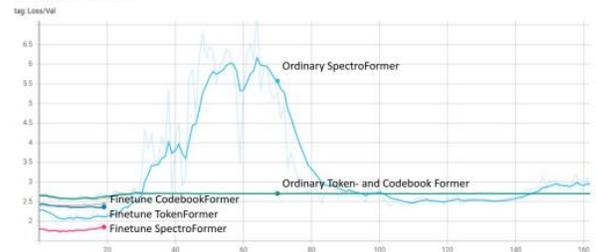

(b) Loss on the validaiton set.

Fig. 3. Execution of every experiment using classification. Numbering the epochs is done on the *x* axis. Since there are no more notable changes after 160 epochs, the *x* scale is eliminated.

### A. Performance Comparison

The optimal outcomes were attained after just 10 training epochs, regardless whether pretrained or not. Only the SpectroFormer models exhibit a noticeable rise over the baseline F1 score of 0.11. Only a little bit over the baseline performance are the optimized Codebook- and TokenFormer. Fig. 4 provides a more thorough perspective of the classification by displaying the confusion matrix for the improved Spectro- and CodebookFormer. It is evident from the confusion matrix that the optimized CodebookFormer is not operating at peak efficiency. It's interesting to see that on the train and validation datasets, all common setups generate comparable results. The training data is neither overfit nor underfit by them. On the other hand, CodebookFormer tends to somewhat underfit, while the best-performing model, the improved SpectroFormer, tends to overfit greatly (refer to Fig. 5).

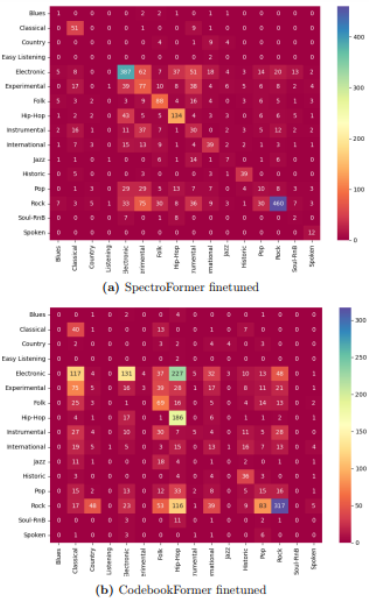

Fig. 4. Matrix of confusion for the validation dataset. The anticipated genre is shown by the row at the bottom, while the ground reality is represented by the left column.

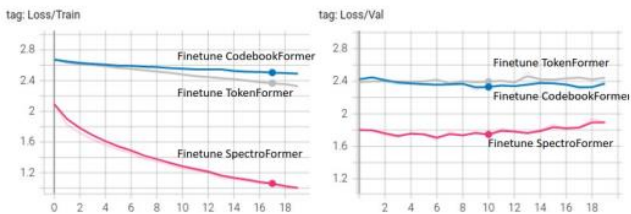

Fig. 5. Loss comparison for the refined models between the train and validation sets

### B. Pretraining Performance

The pretraining loss for each of the three models on the validation set is displayed in Fig. 6. The pretraining train and validation loss do not differ considerably for any of the three models. During pretraining, the models are neither overfit nor underfit. The pretrained SpectroFormer continues to have a larger pretraining loss than the other models across all epochs, while having the best classification performance during finetuning. It should be noted, nevertheless, that while Token- and CodebookFormers' pretraining focuses on correctly classifying tokens, SpectroFormer's pretraining involves regressing Mel spectrogram matrix elements. The pretraining loss of SpectroFormer is not a direct comparison to the other models since they are trained on distinct tasks. However, the identical task is trained on both Codebook- and TokenFormer. It is directly comparable to their pretraining losses. Pretraining results show that CodebookFormer outperforms TokenFormer by 30% when taking into account the loss of the last epoch. Still, after some fine-tuning, CodebookFormer performs almost as well in terms of classification as TokenFormer (see Fig. 3). On the category identification task, the 30% increase in performance from pretraining does not carry over. TokenFormer pretraining loss is reducing till the 300$^{th}$ epoch, though it is not very noticeable on Fig. 6. Also reducing until the 300 epoch is the SpectroFormer loss. As the training comes to a close, the CodebookFormer loss keeps getting lower. Remarkably, in quick shifts, the CodebookFormer loss drops. The atypical behavior observed in the CodebookFormers loss curve may suggest a substantially non-convex (Dirac) loss surface. CodebookFormer's pretraining is redone using an alternate weight initialization due to this unusual loss curve. The repeated experiment loss curve closely resembles the first loss curve.

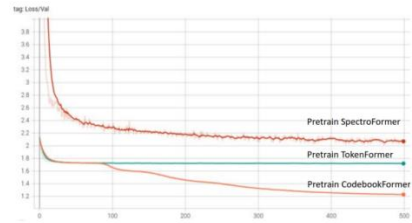

Fig. 6. Pretraining loss in each of the trio of models

### VI. LIMITATIONS

The goal of this study was to investigate the relative efficacy of deep VQ-based audio representations and Mel spectrograms for the identification of musical genres. A definitive response is still tricky due to the empirical nature of this study question and the multiple hyperparameters and design choices that impact it. However, results clearly indicate that Mel spectrograms outperform deep VQ-based representations; the latter are only slightly better than baseline performances, irrespective of the use of tokens or codebooks. Unlike their counterparts based on Fourier transform, deep VQ-based models lack a well-defined and comprehensible structure. Although this non-linearity increases the expressiveness of NNs, it also makes learning more difficult and requires large volumes of data in order to function well. According to [21], who pointed out the substantial data requirements of waveform models like Jukebox, the study holds that the non-linear, data-intensive character of deep VQ-based representations makes genre classification jobs more difficult In this study, Fourier-based representations are superior, although they are not perfect. The phenomenon known as "spectral leakage", which results from the finite length of real-world signals, adds frequencies to the Fourier-transformed signal that were absent from the original. Although window features like the Hann window can lessen its impact, they are unable to completely eradicate it. Spectral leaking is considered a minor issue for MIR applications such as music genre classification. Another result of Fourier operations is phase information, which is usually disregarded in MIR tasks because of its complexity and seeming unpredictability. However, phase information is already taken into consideration by models like as Jukebox, which make music by directly modelling waveforms, negating the need for explicit phase information modelling. This feature highlights the potential usefulness of phase information, as exemplified by the limited yet effective use of the instantaneous phase for harmonic sound creation by GANSynth. The inherent limitations of both Fourier-based and deep VQ-based techniques illustrate the difficult trade-offs required in audio representation for machine intelligence tasks, even while Fourier-based audio representations—particularly Mel spectrograms—demonstrate strong advantages for music genre classification.

## VII. Conclusion and Future Works

Future research on multi-task dialogue systems [22] and audio fingerprinting [23] could greatly enhance music genre classification and interactive communication technologies. The effectiveness of deep VQ audio representations for music genre recognition—a crucial task in MIR—is examined in this study. Because they are more in line with human auditory perception, Fourier-based audio representations have historically been used for MIR tasks; in contrast, deep VQ representations and raw waveforms have proved more popular in scenarios involving music generation. The results, however, support the widely held belief that waveform-based and deep VQ representations might not be the best choices for MIR tasks. Results shows how deep VQ-based models—like TokenFormer and CodebookFormer—perform less effectively in genre recognition because they are unable to adequately capture the subtleties relevant to human hearing. A key point of contention in this study is the amount of data that was utilized for pretraining. Pretraining TokenFormer and CodebookFormer on a dataset 70 times smaller than Jukebox suggests that the amount of training data can have a big impact on model performance. Therefore, SOTA suggests that Jukebox's encoder, which was trained on a significantly larger dataset, is capable of efficiently capturing musical elements pertinent to MIR tasks.

## VIII. Declarations

*A. Funding:* No funds, grants, or other support was received.

*B. Conflict of Interest:* The authors declare that they have no known competing for financial interests or personal relationships that could have appeared to influence the work reported in this paper.

*C. Data Availability:* Data will be made on reasonable request.

*D. Code Availability:* Code will be made on reasonable request.